\documentclass[structabstract]{aa}  
\usepackage{graphicx}
\usepackage{lscape}
\usepackage{txfonts}
\begin{document}
   \title{High angular resolution observations towards OMC-2~FIR~4: Dissecting an intermediate-mass protocluster}\thanks{Based on observations carried out with the IRAM Plateau de Bure Interferometer. IRAM is supported by INSU/CNRS (France), MPG (Germany) and IGN (Spain).}


   \author{A. L\'opez-Sepulcre
          \inst{1}
          \and
          V. Taquet\inst{1}
          \and
          \'A. S\'anchez-Monge\inst{2} 
\and
          C. Ceccarelli\inst{1}
          \and
          C. Dominik\inst{3,}\inst{4}
          \and
          M. Kama\inst{3}
\and
          E. Caux\inst{5,}\inst{6}
\and
          F. Fontani\inst{2} 
\and
          A. Fuente\inst{7}  
\and
          P.T.P. Ho\inst{8,9}       
\and
          R. Neri\inst{10} 
\and
          Y. Shimajiri\inst{11}              
          }

   \institute{UJF-Grenoble 1 / CNRS-INSU, Institut de Plan\'{e}tologie et d\textquoteright Astrophysique de Grenoble (IPAG) UMR 5274, Grenoble, F-38041, France
              \email{ana.sepulcre@obs.ujf-grenoble.fr}
         \and
            Osservatorio Astrofisico di Arcetri, Largo E. Fermi 5, I-50124 Firenze, Italy 
         \and
             Astronomical Institute Anton Pannekoek, University of Amsterdam, Amsterdam, The Netherlands
         \and
             Department of Astrophysics/IMAPP, Radboud University Nijmegen, Nijmegen, The Netherlands   
         \and
             Universit\'e de Toulouse, UPS-OMP, IRAP, Toulouse, France
         \and
             CNRS, IRAP, 9 Av. colonel Roche, BP 44346, 31028 Toulouse Cedex 4, France
         \and
              Observatorio Astron\'omico Nacional, P.O. Box 112, 28803 Alcal\'a de Henares, Madrid, Spain
         \and
            Institute of Astronomy and Astrophysics, Academia Sinica, P.O. Box 23-141, Taipei 106, Taiwan
         \and
            Harvard-Smithsonian Center for Astrophysics, 60 Garden Street, Cambridge, MA 02138, USA   
         \and
             IRAM, 300 rue de la piscine, F-38406 Saint-Martin d'H\`eres, France          
         \and
             Nobeyama Radio Observatory, 462-2 Nobeyama, Minamimaki, Minamisaku, Nagano 384-1305, Japan         
             }
             
   \date{Received; accepted}

 
  \abstract
   {Intermediate-mass stars are an important ingredient of our Galaxy and key to understand how high- and low-mass stars form in clusters. One of the closest known young intermediate-mass protoclusters is OMC-2~FIR~4, located at a distance of 420~pc, in Orion. This region is one of the few where the complete 500-2000 GHz spectrum has been observed with the heterodyne spectrometer HIFI on board the \textit{Herschel} satellite, and unbiased spectral surveys at 0.8, 1, 2 and 3 mm have been obtained with the JCMT and IRAM~30-m telescopes.}
   {We aim to disentangle the core multiplicity and to investigate the morphology of this region in order to study the formation of a low- and intermediate-mass protostar cluster, and aid the interpretation of the single-dish line profiles already in our hands.}
   {We used the IRAM Plateau de Bure Interferometer to image OMC-2~FIR~4 in the 2-mm continuum emission, as well as in DCO$^+$(2--1), DCN(2--1), C$^{34}$S(3--2), and several CH$_3$OH lines. In addition, we analysed observations of the NH$_3$(1,1) and (2,2) inversion transitions made with the Very Large Array of the NRAO. The resulting maps have an angular resolution which allows us to resolve structures of $5''$, equivalent to $\sim 2000$~AU.}
   {Our observations reveal three spatially resolved sources within OMC-2~FIR~4, of one or several solar masses each, with hints of further unresolved substructure within them. Two of these sources have elongated shapes and are associated with dust continuum emission peaks, thus likely containing at least one molecular core each. One of them also displays radio continuum emission, which may be attributed to a young B3-B4 star that dominates the overall luminosity output of the region. The third source identified displays a DCO$^+$(2--1) emission peak, and weak dust continuum emission. Its higher abundance of DCO$^+$ relative to the other two regions suggests a lower temperature and therefore its possible association with either a younger low-mass protostar or a starless core. It may alternatively be part of the colder envelope of OMC-2~FIR~4.}
   {Our interferometric observations evidence the complexity of the intermediate-mass protocluster OMC-2~FIR~4, where multiple cores, chemical differentiation, and an ionised region all coexist within an area of only 10000~AU.}

   \keywords{ISM: individual objects: OMC-2~FIR~4 --
                ISM: molecules
               }
\titlerunning{Dissecting an intermediate-mass protocluster}
   \maketitle
%

\section{Introduction}

Star formation starts in dense fragments within clouds of molecular gas and dust and ends when nuclear hydrogen burning begins at the centre of a newly formed star. During this evolution, the physical processes and different phases involved are not necessarily the same for stars of all masses. Indeed, in the field of star formation there has traditionally been a division between low-mass stars ($M \lesssim 2$~M$_\odot$), intermediate-mass stars (2~M$_\odot < M \lesssim 8$~M$_\odot$), and high-mass stars ($M > 8$~M$_\odot$).

In the past decades, most of the attention has been devoted to the study of low-mass and high-mass star formation, revealing differences between the two which are not only dynamical (Beuther et al.~\cite{beu07}, Zinnecker \& Yorke~\cite{zin07}), but also chemical in nature (Bottinelli ~\cite{bot06}, Ceccarelli et al.~\cite{cec07}). The formation of intermediate-mass (IM) stars represents the bridge between low-mass and high-mass star formation and can bring important pieces of information to understand how differently the process of star formation works in the two mass extremes. As their high-mass counterparts, IM stars emit a non negligible amount of radiation in the far ultraviolet (FUV), and are thus able to photodissociate and ionise their surroundings. Therefore, the study of IM star formation can aid the current debate on the formation mechanism of high-mass stars (see Zinnecker \& Yorke~\cite{zin07} for a review).

While low-mass stars may form in isolation or in loose stellar aggregates, high-mass stars are always seen to be formed in densely clustered environments. IM star formation also occurs in clustered mode (e.g. Fuente et al.~\cite{asun07}) and has been found to be the transition between loose low-mass aggregations and tight high-mass clusters (Testi et al.~\cite{testi99}). IM protoclusters represent excellent laboratories to study clustered star formation at almost the full range of stellar masses. Observationally, they can typically be found closer ($d \lesssim 1$~kpc) than high-mass protoclusters, and they are also less complex. 

Despite all this, relatively little work has focused on the early stages of IM star formation. Only a few individual sources have been studied in detail (e.g. Fuente at al.~\cite{asun09}, Crimier et al.~\cite{crim09}, \cite{crim10}, S\'anchez-Monge et al.~\cite{alvaro10}, Palau et al.~\cite{aina11}, van Kempen et al.~\cite{vk12}), and therefore much remains unknown about the formation and first evolutionary stages of these objects.

The present work focuses on the IM protocluster OMC-2~FIR~4, with a luminosity that has been reported to be between 50~L$_\odot$ (Adams et al.~\cite{adams12}) and 1000~L$_\odot$ (Crimier et al.~\cite{crim09}). It is located in the Orion~A complex, 2~pc North of the famous Orion Nebula (M42) and the Trapezium OB association, and lies $\sim$420~pc away from the sun (Menten et al.~\cite{menten07}, Kim et al.~\cite{kim08}).

Due to its relative proximity, multiple studies have been carried out towards this region and the OMC-2/3 filament where it is located. These include continuum observations at several wavelengths (e.g. Chini et al.~\cite{chini97}, Johnstone \& Bally~\cite{jb99}, Nielbock et al.~\cite{ni03}), and various molecular line studies (e.g. Tatematsu~\cite{tate93}, Cesaroni \& Wilson~\cite{cesa94}, Takahashi et al.~\cite{taka08}, Tatematsu et al.~\cite{tate08}, Liu et al.~\cite{liu11}, Li et al.~\cite{li12}). Most of them, however, are coarse angular-resolution observations, very useful to investigate the overall structure of the whole filament, but not sufficient to resolve the substructure within the OMC-2~FIR~4 region. The only dedicated interferometric observations towards this source are Very Large Array (VLA) maps carried out at 3.6~cm by Reipurth et al.~(\cite{rei99}), and a recent study conducted with the Nobeyama Millimeter Array (NMA) at 3~mm by Shimajiri et al.~(\cite{shima08}; hereafter S08). The latter reveal several cores and the authors speculate that an external outflow driven by the nearby FIR~3 is responsible for triggering fragmentation in FIR~4, highlighting the complexity of this region.

OMC-2~FIR~4 is also one of the targets of the  \textit{Herschel}\footnote{\textit{Herschel} is a European Space Agency (ESA) space observatory with science instruments provided by European-led principal investigator consortia and with important participation from the National Aeronautics and Space Administration (NASA).} Guaranteed Time Key Programme Chemical HErschel Surveys of Star forming regions (CHESS\footnote{http://www-laog.obs.ujf-grenoble.fr/heberges/chess/};  Ceccarelli et al.~\cite{cec10}, Kama et al.~\cite{kama12}), which has conducted unbiased spectral surveys with the HIFI spectrometer (de Graauw et al.~\cite{graauw10}) of eight star forming regions, each representative of a particular aspect of the star formation process: evolutionary stage, mass of the forming star, and/or interaction with the surroundings. The HIFI half-power beam width ranges between 11$''$ at 1900~GHz and 41$''$ at 500~GHz, thus covering most, if not all, the area of OMC-2~FIR~4. High-angular resolution observations are therefore a crucial tool to help disentangle the information hidden in the HIFI line profiles.

With this in mind, we report in this paper 2-mm continuum and line observations performed with the IRAM Plateau de Bure Interferometer (PdBI) and ammonia maps obtained with the VLA towards OMC-2~FIR~4. We describe these observations in Sect.~\ref{obs}; in Sect.~\ref{results} we present the maps and spectra obtained, which we discuss and interpret in Sect.~\ref{discuss}. Finally, Sect.~\ref{conclusions} summarises the main conclusions of this work.


\section{Observations}\label{obs}

\subsection{Plateau de Bure observations}

We observed OMC-2~FIR~4 with the IRAM Plateau de Bure Interferometer (PdBI) in two tracks on 15 October 2010 and 15 April 2011, using the most compact D and C configurations of the array. The data presented in this paper cover a spectral window of 1.8~GHz obtained with the WIDEX correlator, whose channel spacing is 1.95~MHz ($\sim$4~km~s$^{-1}$), centred at a frequency of 143.4~GHz. The phase centre of the observations is $\alpha$(J2000)~=~$05^\mathrm{h}35^\mathrm{m}26.971^\mathrm{s}$, $\delta$(J2000)~=~--05$^\circ09'56.77''$. The systemic velocity of OMC-2~FIR~4 is $V_\mathrm{LSR} = 11.4$~km~s$^{-1}$.

The bandpass of the receivers was calibrated by observing 0923+392 in Oct 2010 and NRAO150 in Apr 2011, with a flux density of 3.55~Jy and 2.53~Jy, respectively. Amplitude and phase calibrations were achieved by monitoring 0430+052 and 0528+134, whose flux densities were determined to be 1.32 and 1.67~Jy, respectively, at 143.4~GHz. The uncertainty in the amplitude calibration is estimated to be  30\%.

The data were calibrated and analysed with the GILDAS\footnote{The GILDAS package is available at http://www.iram.fr/IRAMFR/GILDAS} software package developed at IRAM and the Observatoire de Grenoble. We obtained continuum maps from the line free channels and subtracted the continuum from the line emission directly in the (u,v)-domain. The resulting 1$\sigma$ RMS and beam size for the naturally weighted continuum map are 4.9~mJy~beam$^{-1}$ and $4\farcs87 \times 2\farcs73$, respectively (see Table~\ref{tcont}). The maps presented in this paper are not corrected for the primary beam, but we did apply the correction to measure the reported fluxes.

\subsection{Very Large Array ammonia observations}\label{vlaobs}

The Very Large Array (VLA\footnote{The Very Large Array (VLA) is operated by the
National Radio Astronomy Observatory (NRAO), a facility of the National Science
Foundation operated under cooperative agreement by Associated Universities,
Inc.}) was used to observe the ammonia ($J$,$K$) = (1,1) and (2,2) inversion
transitions simultaneously (project AC556). Twenty adjacent pointing centres,
covering OMC-2 and OMC-3 regions, were carried out on 29 July 2000 and 24
September 2000, with the array in the D (compact) configuration. The FWHM of the
primary beam at the frequency of the observations is approximately $2'$; the field
centres were separated by $1'$ to ensure full sampling. Each field was observed
for a total on-source integration time of approximately 30~minutes. The absolute
flux scale was set by observing the quasar 1328+307 (3C286), for which we
adopted a flux of 2.5~Jy. The quasars 0605$-$085 and 0539$-$057, with
bootstrapped fluxes of 2.4~Jy and 1.0~Jy, respectively, were observed regularly
to calibrate the gains and phases. Bandpass calibration was performed by
observing the bright quasar 0316+413 (3C84) with a flux density of 14.2~Jy (for
July) and 17.5~Jy (for September).

\begin{table*}[!hbt]
\centering
\caption{Parameters of our continuum observations and those of Shimajiri et al. (\cite{shima08})}
\begin{tabular}{lcccc}
\hline
Parameter & \multicolumn{2}{c}{This work} & \multicolumn{2}{c}{Shimajiri et al.~(\cite{shima08})} \\
\hline
Interferometer & PdBI & PdBI & NMA & NMA\\
Frequency (GHz) & 143.4 & 143.4 & 92 & 92\\
Weighting & natural & uniform & natural & uniform\\
Primary Beam ($''$) & 35 & 35 & 82 & 82 \\
Baseline range (k$\lambda$) & 10 - 80 & 10 - 80 & 2.9 - 115 & 2.9 - 115\\
Beam size (arcsec) & 4.87$\times$2.73 & 3.74$\times$2.74 & 6.96$\times$4.03 & 6.55$\times$3.34\\
P.A. of the beam ($^\circ$) & --0.76 & --12.98 & --34.59 & --40.31\\
R.M.S. (mJy~beam$^{-1}$) & 4.9 & 4.6 & 1.2 & 1.4\\
\hline
\end{tabular}
\label{tcont}
\end{table*}

The data reduction followed the VLA standard
guidelines for calibration of high-frequency data, using the NRAO package AIPS.
The NH$_3$\,(1,1) and NH$_3$\,(2,2) lines were observed simultaneously in the
four IF correlator modes of the VLA (with two polarisations for each line),
providing 63 channels with a spacing of 0.62~km~s$^{-1}$ across a
bandwidth of 3.125~MHz, plus an additional continuum channel containing the
central 75 per cent of the total bandwidth. The bandwidth was centred at the
systemic velocity $V_\mathrm{LSR}$=11.4~km~s$^{-1}$.

Calibrated uv-data were loaded in the software MIRIAD (Sault et al.\ 1995) to
perform the imaging. While the continuum emission was found to be negligible 
(we measured an upper limit of about 2~mJy at the frequency of the observations)
we subtracted any possible related contribution from line-free channels
using the task \texttt{UVLIN}. We combined and imaged the data from the central
part of the mosaic, covering the OMC-2 region, with the task \texttt{MOSSDI}.
The rms noise in the naturally weighted maps is 5~mJy~beam$^{-1}$ per
0.62~km~s$^{-1}$ channel. 
We applied a \emph{uv}-taper function of
60~k$\lambda$ (3$''$ in the image plane) to improve the signal-to-noise ratio. The
resulting synthesised beam is $6\farcs2\times5\farcs3$, with PA=$-$5$^\circ$.

\section{Results}\label{results}

\subsection{Millimetre continuum emission}\label{cont}

\begin{figure}[!htb]
   \centering
   \includegraphics[scale=0.5]{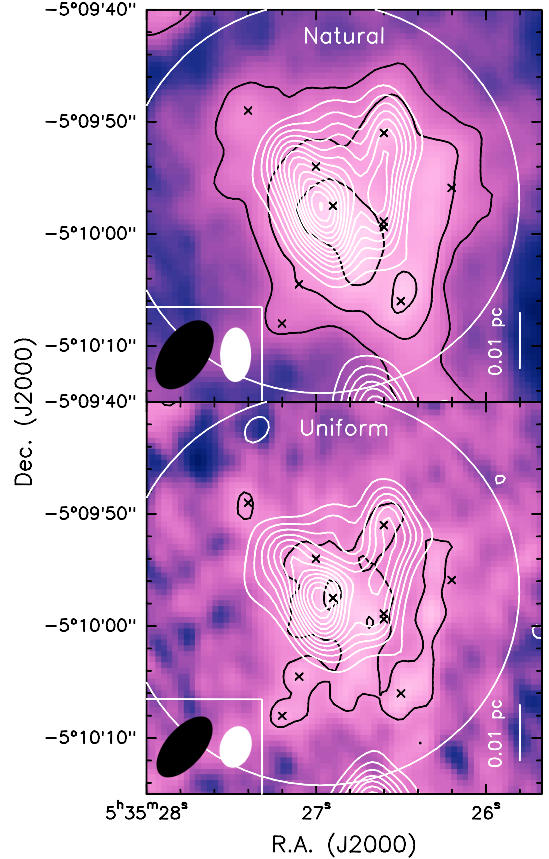}
      \caption{Interferometric continuum maps of OMC-2~FIR~4 with natural (\textit{top}) and uniform weighting (\textit{bottom}). White contours represent the 2-mm continuum emission obtained with the PdBI, while black contours, as well as the background colour scale, correspond to the 3-mm continuum emission seen with the NMA (S08). Contours for both the PdBI and NMA maps start at 3$\sigma$ and increase by steps of 3$\sigma$. The bottom-left ellipses represent the respective beam sizes. The white open circles mark the primary beam $FWHM$ for the PdBI observations. The black crosses mark the NMA 3-mm cores as reported in Table~4 of S08.}
         \label{fcont}
   \end{figure}

Figure~\ref{fcont} shows our PdBI naturally and uniformly weighted 2-mm continuum maps (white contours). For comparison, these are overlaid on the 3-mm continuum images obtained by S08 with the Nobeyama Millimeter Array (NMA). From their uniformly weighted image, S08 identify 11 cores at 3~mm within OMC-2~FIR~4, whose positions according to their Table~4 are marked with black crosses in the figure. The parameters of our continuum observations and those of S08 are listed in Table~\ref{tcont}.

The PdBI images allow us to resolve two elongated structures in OMC-2~FIR~4, separated by $\sim 5''$, surrounded by more extended emission which covers a total size of about 15$''$. The uniform map suggests the presence of additional fragmentation which cannot be completely resolved at the angular resolution of the observations. On the other hand, the NMA images display fainter, more complex and more extended emission, which is largely below the 3$\sigma$ level (first contour) in the case of the uniform map.

\begin{table*}[!bht]
\caption{Lines observed with the PdBI and VLA interferometers} 
\label{tlines}     
\centering  
\begin{tabular}{lccccc}
\hline\hline 
Line & $\nu$\tablefootmark{a} & $E_\mathrm{up}$ & $A_{ul}$ & 1$\sigma$ RMS\tablefootmark{b} \\ 
 & (GHz) & (K) & (s$^{-1}$) & (mJy~beam$^{-1}$)\\ 
\hline   
CH$_3$OH($3_{1,3} - 2_{1,2}$~A$^+$) & 143.865801 & 28.3 & 1.07$\times10^{-5}$ & 3.7\\
CH$_3$OH($3_{0,3} - 2_{0,2}$~E) & 145.093707 & 27.1 & 1.23$\times10^{-5}$ & ---\\
CH$_3$OH($3_{-1,3} - 2_{-1,2}$~E) & 145.097370 & 19.5 & 1.10$\times10^{-5}$ & ---\\
CH$_3$OH($3_{0,3} - 2_{0,2}$~A$^+$) & 145.103152 & 13.9 & 1.23$\times10^{-5}$ & ---\\
CH$_3$OH($3_{2,2} - 2_{2,1}$~A$^-$) & 145.124410 & 51.6 & 6.89$\times10^{-6}$ & ---\\
CH$_3$OH($3_{2,1} - 2_{2,0}$~E) & 145.126190 & 36.2 & 6.77$\times10^{-6}$ & ---\\
CH$_3$OH($3_{-2,2} - 2_{-2,1}$~E) & 145.126392 & 39.8 & 6.85$\times10^{-6}$ & ---\\
CH$_3$OH($3_{1,2} - 2_{1,1}$~E) & 145.131855 & 35.0 & 1.12$\times10^{-5}$ & ---\\
CH$_3$OH($3_{2,1} - 2_{2,0}$~A$^+$) & 145.133460 & 51.6 & 6.89$\times10^{-6}$ & ---\\
\hline
DCO$^+$(2 -- 1)\tablefootmark{c} & 144.077289 & 10.4 & 2.12$\times10^{-4}$ & 3.6\\
\hline
C$^{34}$S(3 -- 2) & 144.617101 & 13.9 & 5.78$\times10^{-5}$ & 3.2\\
\hline
DCN(2 -- 1)\tablefootmark{c} & 144.828002 & 10.4 & 1.27$\times10^{-4}$ & 3.4\\
\hline  
NH$_3$(1,1) & 23.69450 & 23.3 & 1.68$\times10^{-7}$ & 5.0\\
NH$_3$(2,2) & 23.72263 & 64.4 & 2.24$\times10^{-7}$ & 4.7\\
\hline
\end{tabular}
\\
\tablefoottext{a}{Millimetre line frequencies from the CDMS catalogue: http://www.astro.uni-koeln.de/cdms/}\\
\tablefoottext{b}{--- indicates no map is presented in this paper}\\
\tablefoottext{c}{Transition with 6 hyperfine components}
\end{table*}

It is worth mentioning that the primary beam size of our PdBI observations is 35$''$, considerably smaller than that of the NMA maps, which is 82$''$. Similarly, the shortest baseline attained with the NMA is 2.9~k$\lambda$ while that of the PdBI is 10~k$\lambda$. This results in a higher sensitivity of the NMA to extended structures. More specifically, our shortest baseline implies that our observations are sensitive to structures smaller than 17$''$ down to a level of 10\% (see Eq.~A8 in Wilner \& Welch~\cite{ww94}, or Eq. A6 in Palau et al.~\cite{aina10}), while for the NMA images this number is as high as 56$''$. Thus, we are not sensitive to the cores identified by S08 that fall outside the outermost PdBI contour. In spite of this, we still have several concerns regarding the core identification made by S08. First, it was based on the uniform map, which has a low signal-to-noise (S/N) ratio; second, the coordinates given for some of the cores do not coincide with the corresponding emission peak positions in the image; and third, two pairs of cores lie within a beam size of each other. We believe the NMA natural map is more reliable than the uniform one given its higher sensitivity. This already shows a number of clearly separated cores, some of which appear to be followed also by the PdBI contours. In conclusion, even though we prefer to be cautious about the reality of the 11 cores reported in S08, we are convinced that OMC-2~FIR~4 is composed of several cores. Based on our PdBI maps, we will hereafter refer to the dominant eastern region as \textit{main} source, and to the western one as \textit{west} source.

The continuum fluxes, measured inside the 5$\sigma$ contour of the corresponding maps, are $1.06 \pm 0.32$~Jy and $1.10 \pm 0.33$~Jy for the natural and uniform maps, respectively. We can estimate the mass of the emitting clump by considering that all the flux is emitted by dust. This is a reasonable assumption, since Reipurth et al.~(\cite{rei99}) reported a 3.6-cm flux of 0.64~mJy for OMC-2~FIR~4 (see Sect.~\ref{hii}), and therefore the contribution of free-free emission at 2~mm is negligible, i.e. about three orders of magnitude below the total flux we measure. Assuming that the dust emission is optically thin, the mass, $M_\mathrm{dust}$ is given by

\begin{equation}
M_\mathrm{dust} = \frac{S_\nu\ d^2}{\kappa_\nu\ B_\nu(T_\mathrm{d})\ R_\mathrm{d}}
\label{emdust}
\end{equation}
where $S_\nu$ is the flux we meaure inside the 5$\sigma$ contour at 2~mm, $d$ the distance to OMC-2~FIR~4, $T_\mathrm{d}$ the dust temperature (adopted to be 50~K from S08), $B_\nu(T_\mathrm{d})$ is the Planck black body function for a temperature $T_\mathrm{d}$, and $R_\mathrm{d}$ is the dust-to-gas ratio, assumed equal to 0.01. $\kappa_\nu = \kappa_0 (\frac{\nu}{\nu_0})^\beta$ is the frequency-dependent dust mass opacity coefficient. Chini et al.~(\cite{chini97}) derived $\beta = 2$ from the Spectral Energy Distribution fit towards OMC-2~FIR~1 and FIR~2. Since FIR~4 is in the same molecular filament as FIR~1 and FIR~2, we have adopted $\beta = 2$, and $\kappa_0 = 1$~cm$^2$~g$^{-1}$ at $\nu_0 = 250$~GHz (Ossenkopf \& Henning~\cite{oh94}).

The resulting mass, from the naturally-weighted 2-mm continuum map, is $9 \pm 3$~M$_\odot$. This should be considered as a lower limit to the total mass of OMC-2~FIR~4 (dense cores plus envelope), since a percentage of the total flux is expected to be filtered out by the interferometer. We can estimate the fraction of flux lost with the aid of the work published by Crimier et al.~(\cite{crim09}), who modelled the Spectral Energy Distribution (SED) function observed towards OMC-2~FIR~4 at infrared and sub-millimetre wavelengths. The flux they predict at 2~mm is 2~Jy, implying a flux loss of 45\% in our PdBI continuum observations.

Crimier et al.~(\cite{crim09}) derived a mass of 30~M$_\odot$ for the envelope of OMC-2~FIR~4, i.e. higher than our estimate by a factor 3. While this difference may be partly due to the loss of flux in our interferometric maps, we note that a different choice of dust temperature in Eq.~\ref{emdust} can significantly change the obtained mass. For example, adopting $T_\mathrm{d} = 20$~K, which is a reasonable dust temperature for the source outer envelope (Lis et al.~\cite{lis98}, J\o rgensen et al.~\cite{jor06}, Crimier et al.~\cite{crim09}), yields a mass of 26~M$_\odot$.

\subsection{Line emission}\label{line}

The list of detected transitions are presented in Table~\ref{tlines}, together with the 1$\sigma$~RMS values of the corresponding velocity channel maps (see Sects.~\ref{pdb} and \ref{nh3}). Out of the nine methanol lines detected, eight are blended due to the poor spectral resolution of the observations. Therefore, the methanol maps shown in this paper correspond to the only isolated transition we observe, i.e. CH$_3$OH(3,1,+0 -- 2,1,+0) at 143.87~GHz, whose upper level energy is 28~K and which will hereafter be referred to as CH$_3$OH for simplicity.

   \begin{figure}[!htb]
   \centering
   \includegraphics[angle=-90,scale=0.5]{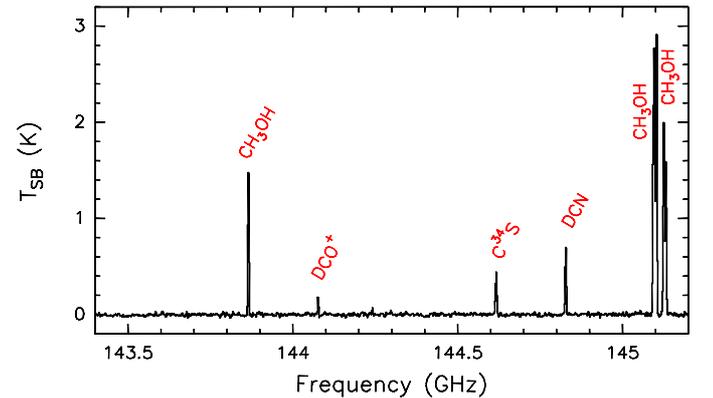}
      \caption{PdBI spectrum obtained towards the phase centre of OMC-2~FIR~4 with the WIDEX correlator. The species of the detected lines are labelled.}
         \label{fspt}
   \end{figure}

\subsubsection{PdBI molecular line spectra and maps}\label{pdb}

\begin{figure*}[!bht]
 \centering
 \includegraphics[angle=-90,scale=0.65]{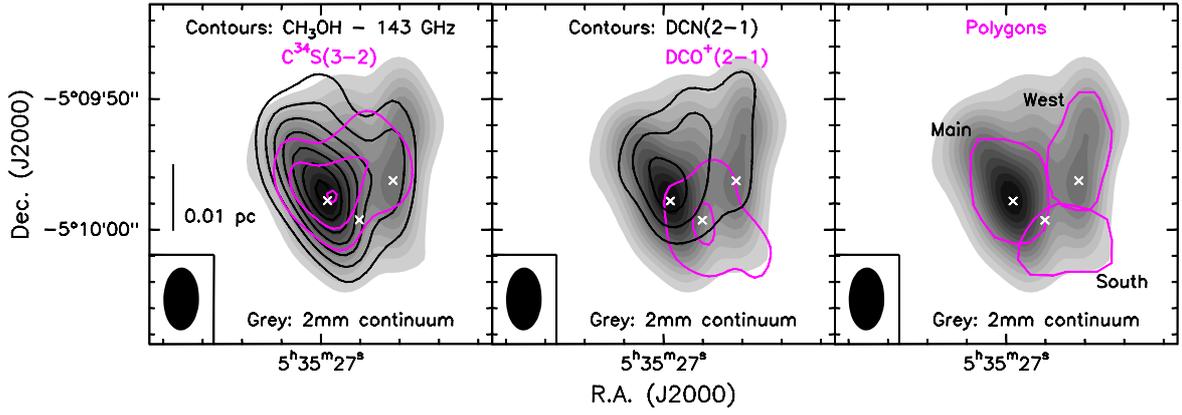}
 \caption{Velocity-integrated PdBI contour maps obtained towards OMC-2~FIR~4, overlaid on the naturally weighted continuum map (grey scale). The bottom-left ellipse in each panel represents the beam size. \textit{Left}: CH$_3$OH(3,1,+0 -- 2,1,+0) (black) and C$^{34}$S(3--2) (magenta). \textit{Centre}: DCN(2--1) (black) and DCO$^+$(2--1) (magenta). All the contours start at 3$\sigma$ and increase by steps of 3$\sigma$. The 1$\sigma$ RMS values are 0.16, 0.14, 0.17, and 0.08~Jy~beam$^{-1}$~km~s$^{-1}$, respectively for the CH$_3$OH, C$^{34}$S(3--2), DCN(2--1), and DCO$^{+}$(2--1) integrated maps. \textit{Right}: Polygons representing the three differentiated components, \textit{main}, \textit{west}, and \textit{south} (see text), used to extract average spectra. The white crosses mark the positions of the three sources identified in this work (see Table~\ref{tflux}).}
 \label{fint}
 \end{figure*}

The 1.8-GHz wide spectrum obtained at the peak coordinates of OMC-2~FIR~4 with the WIDEX correlator at PdBI is presented in Fig.~\ref{fspt}, where the detected species are labelled. Since we are also in possession of IRAM~30-m single-pointing data towards the same source (L\'opez-Sepulcre et al., \textit{in prep.}), we provide in Table~\ref{tloss} an estimate of the velocity-integrated flux loss for each PdBI line. This ranges from 13\% for CH$_3$OH to 73\% for DCO$^+$(2--1), and indicates the presence of an important contribution from extended emission in these tracers which is filtered out by the interferometer.

 \begin{table}[!h]
\caption{Flux loss for the PdBI molecular lines} 
\label{tloss}     
\centering  
\begin{tabular}{lc}
\hline\hline 
Line & Flux loss \\
 & (\%) \\ 
\hline   
CH$_3$OH & 13 \\
DCO$^+$(2 -- 1) & 73\\
C$^{34}$S(3 -- 2) & 60\\
DCN(2 -- 1) & 72\\
\hline
\end{tabular}
\end{table}
   
\begin{figure*}[!htb]
   \centering
   \begin{tabular}{c}
   \includegraphics[angle=-90,scale=0.65]{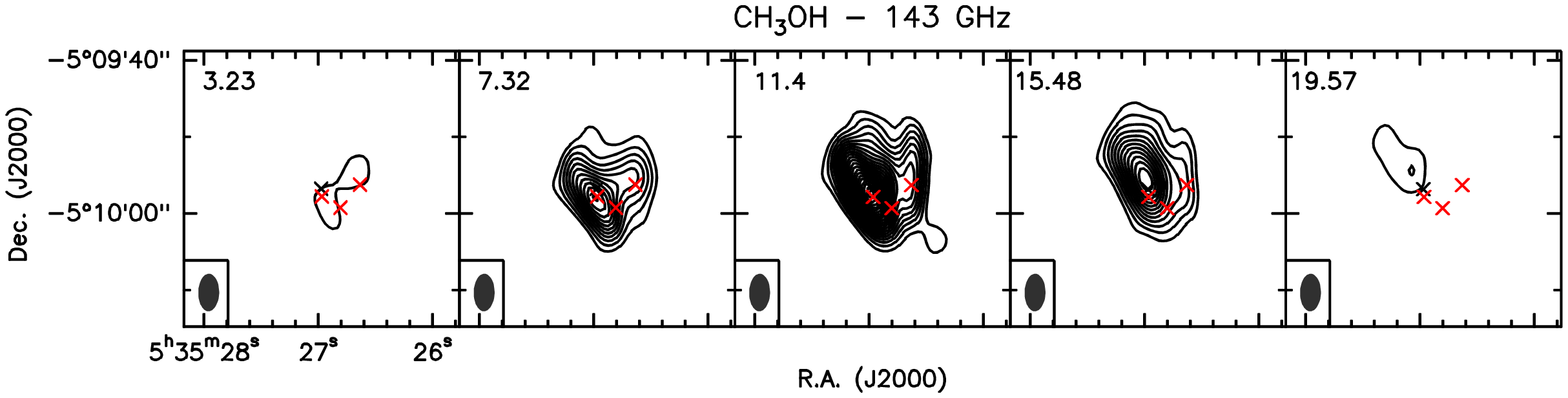}\\
   \includegraphics[angle=-90,scale=0.65]{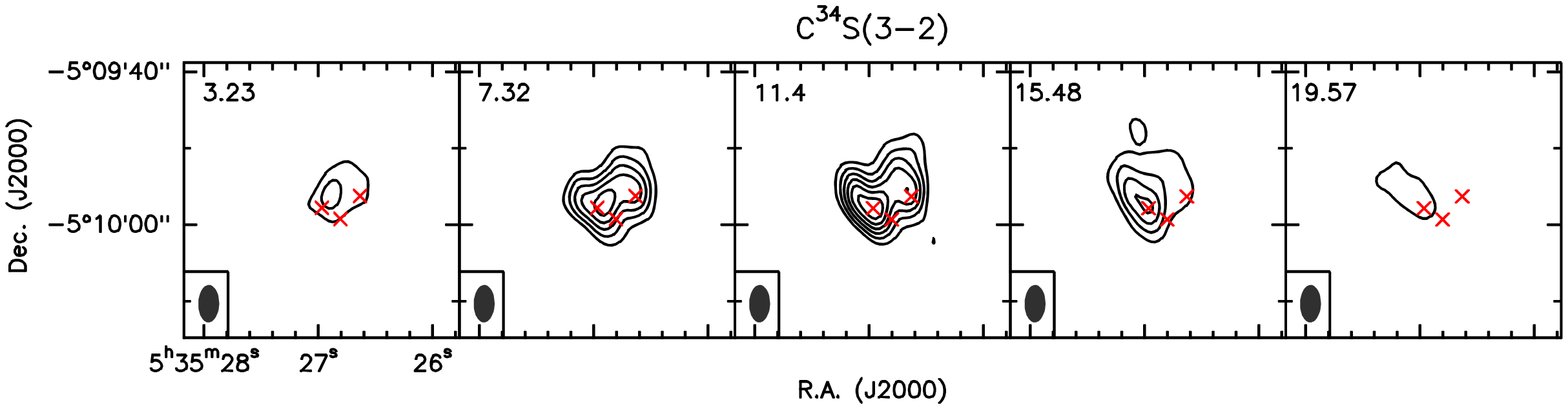}\\
   \includegraphics[angle=-90,scale=0.65]{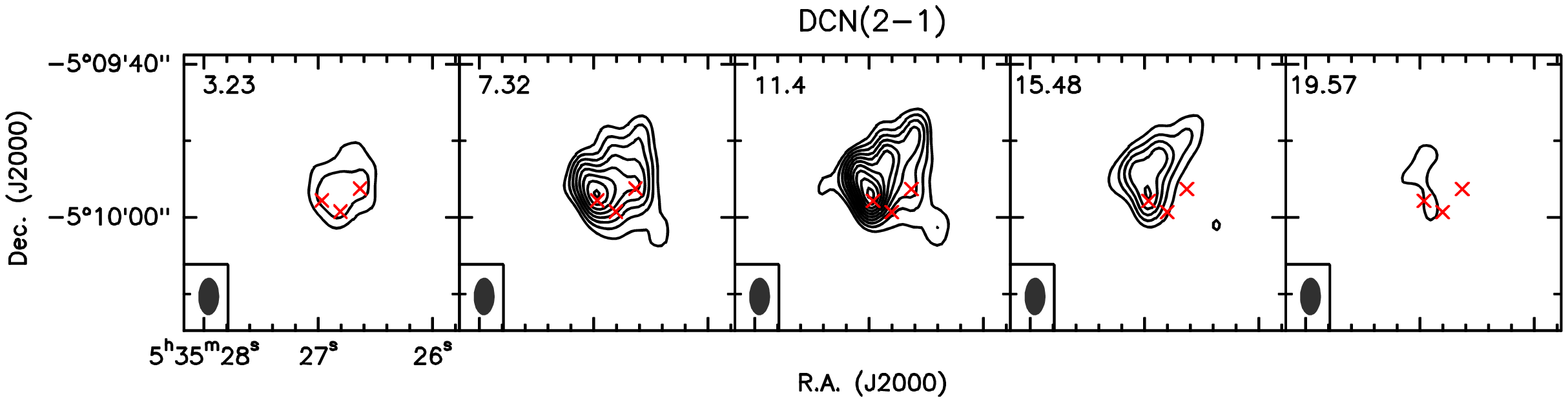}\\ 
   \includegraphics[angle=-90,scale=0.65]{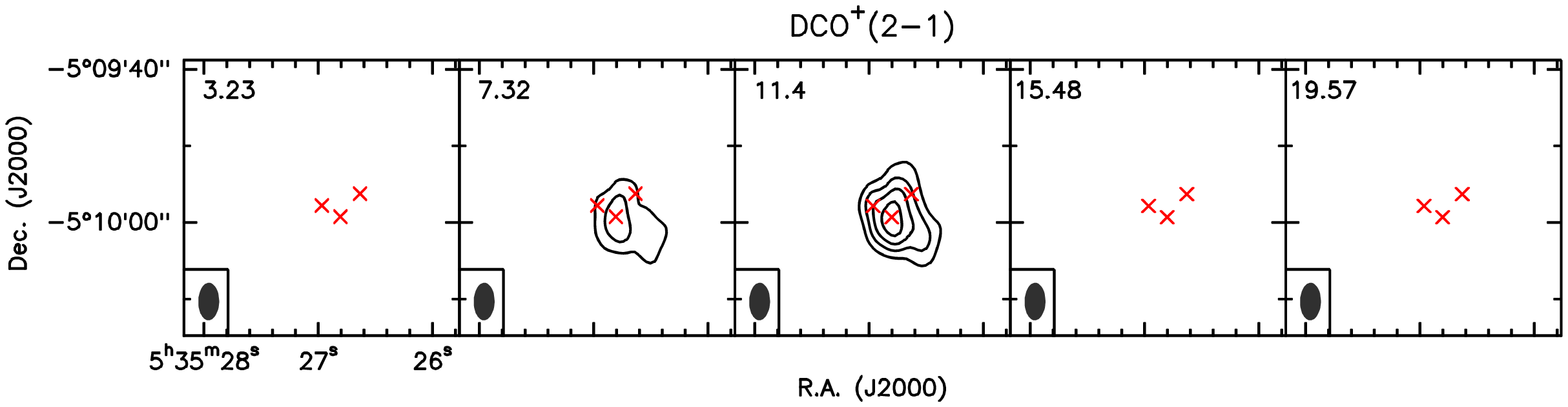}\\   
   \end{tabular}
      \caption{Channel contour maps of each line detected with PdBI towards OMC-2~FIR~4. Contours start at 5$\sigma$ and increase by steps of 5$\sigma$. The 1$\sigma$ RMS values are listed in Table~\ref{tlines}. The velocity of each channel is marked on the top-left corner of each panel. Black crosses mark the phase centre of the observations, i.e. $\alpha$(J2000)~=~05$^\mathrm{h}$35$^\mathrm{m}$26.971$^\mathrm{s}$, $\delta$(J2000)~=~--05$^\circ$09$'$56.77$''$, while red crosses mark the positions of the three sources identified in this work (see Table~\ref{tflux}). The bottom-left ellipses in each panel represent the beam sizes.}
         \label{fchan}
   \end{figure*}

Given the low spectral resolution of the WIDEX spectrum (1.95~MHz, i.e. 4~km~s$^{-1}$), it is not possible to analyse the shape of the line profiles. However, this is beyond the scope of the present work, which aims primarily at investigating the spatial distribution of the emission in the different tracers. To this end we provide, for each millimetre molecular tracer, velocity-integrated maps in Fig.~\ref{fint}, and velocity channel maps in Fig.~\ref{fchan}. The synthesised beam size of these images is 4.8$\times$2.7 arcsec, and is depicted on the bottom-left corner of each panel in the figures.

\begin{table*}[!bht]
\caption{OMC-2~FIR~4 sources: coordinates and fractional fluxes for each tracer} 
\label{tflux}     
\centering  
\begin{tabular}{lccccccc}
\hline\hline 
Source & R.A.(J2000) & Dec.(J2000) & CH$_3$OH & C$^{34}$S(3--2) & DCN(2--1) & DCO$^{+}$(2--1) & Cont. (natural) \\ 
\hline   
Total\tablefootmark{a} &  &  & 13.0 & 3.7 & 6.3 & 1.2 & 1.1 \\
\hline
Main(\%) & $05^\mathrm{h}35^\mathrm{m}26.97^\mathrm{s}$ & $-05^\circ09'57.8''$ & 49 & 51 & 61 & 39 & 35 \\
West(\%) & $05^\mathrm{h}35^\mathrm{m}26.63^\mathrm{s}$ & $-05^\circ09'56.2''$ & 15 & 25 & 30 & 32 & 18 \\
South(\%) & $05^\mathrm{h}35^\mathrm{m}26.80^\mathrm{s}$ & $-05^\circ09'59.3''$ & 13 & 11 & 12 & 65 & 12\\
\hline  
\end{tabular}\\
\tablefoottext{a}{Measured within the 5$\sigma$ and 3$\sigma$ contour for the continuum and line maps, respectively; flux units in Jy for continuum and Jy~km~s$^{-1}$ for lines}
\end{table*}

If we focus first on the integrated maps (Fig.~\ref{fint}), the morphological complexity of the source becomes evident from a comparison of the emission distribution in the different tracers:

\begin{itemize}
\item[-] The CH$_3$OH emission is dominated by the \textit{main} source (see Sect.~\ref{cont} above), which displays an elongated shape, and shows hints of a secondary component towards the west (the above mentioned \textit{west} source), mimicking the spatial distribution of the 2-mm continuum emission. This morphology does not change when seen in the blended CH$_3$OH lines around 145~GHz.
\item[-] C$^{34}$S(3--2) displays a more compact emission that traces both the \textit{main} and \textit{west} sources.
\item[-] The DCN(2--1) emission peaks at the \textit{main} source, but it expands towards the west and the north, following the elongated morphology of the \textit{west} source as seen in the continuum images.
\item[-] DCO$^{+}$(2--1) peaks to the South of the \textit{main} and \textit{west} sources, a different region which we name \textit{south} source, not well traced by any other line.
\end{itemize}

In summary, from the integrated maps we conclude that each molecular line traces the region differently. We differentiate three spatially separated emission components, which we have named \textit{main}, \textit{west}, and \textit{south} sources, lying $\sim 0.01$~pc (i.e. 2000~AU) apart from each other. These regions are schematically depicted in the right panel of Fig.~\ref{fint} by three magenta polygons that have been manually defined to pinpoint similar central areas for each while keeping approximately their morphological shapes. From these we have extracted the respective continuum and line fluxes for each tracer. The resulting fractional fluxes with respect to the total flux, measured inside the $3\sigma$ map contour for the lines and the $5\sigma$ contour for the continuum map, are listed in Table~\ref{tflux}. The central coordinates of each source correspond to the emission peaks of the 2-mm continuum image for the \textit{main} and \textit{west} sources, and those of the DCO$^+$(2--1) map for the \textit{south} one. Notice that, with the exception of DCO$^{+}$(2--1), due to the relatively small area of each polygon, the sum of fluxes is smaller than the total flux in the whole region. The sum amounts to 136\% in the case of DCO$^+$(2--1) partly because the polygons corresponding to the \textit{main} and \textit{south} region overlap around the position of the peak emission, and partly because there is an amount of contributing flux outside the 3$\sigma$ contour. This table reflects the predominance of the \textit{main} source's flux over that of the other two for all the tracers except DCO$^{+}$(2--1), for which the \textit{south} source dominates the overall emission.

Using Eq.~\ref{emdust}, we have computed the mass, $M$, of each source for two values of the dust temperature: 20 and 50~K. These are listed in Table~\ref{tmass}, together with the corresponding \textbf{source -averaged} gas column densities derived from the equation:

\begin{equation}
N_\mathrm{H_2} = \frac{M}{\mu\ m_\mathrm{H}\ \Omega\ d^2}
\label{ecoldens}
\end{equation}
where $m_\mathrm{H}$ is the mass of the hydrogen atom, $\mu = 2.33$ is the mean molecular mass in units of hydrogen atom masses, and $\Omega$ is the solid angle subtended by the dust continuum emission. Assuming the depth of the region equals the projected one (i.e. 5000~AU), we derive an average density in the range $0.9 - 2.6 \times 10^7$~cm$^{-3}$ for OMC-2~FIR~4.
   
\begin{figure*}[!hbt]
   \centering
   \includegraphics[angle=-90,scale=0.7]{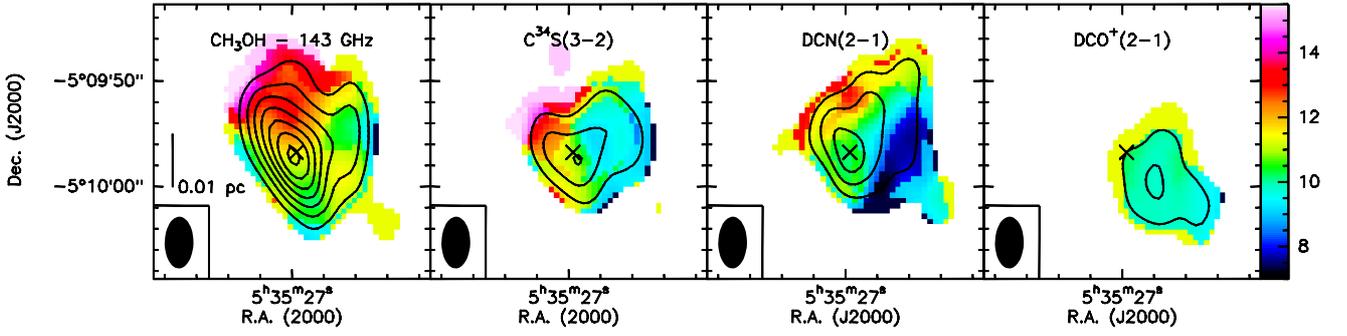}
      \caption{Velocity maps (colour scale, in km~s$^{-1}$) of each line detected with PdBI towards OMC-2~FIR~4, with overlaying contours of the corresponding velocity-integrated maps. Contour levels are as in Fig.~\ref{fint}. The bottom-left ellipse in each panel represents the beam size. Crosses mark the phase centre of the observations, i.e. $\alpha$(J2000)~=~05$^\mathrm{h}$35$^\mathrm{m}$26.971$^\mathrm{s}$, $\delta$(J2000)~=~--05$^\circ$09$'$56.77$''$.}
         \label{fvelo}
   \end{figure*}

The channel maps in Fig.~\ref{fchan} reveal more details about the spatial distribution of the molecular line emission. Despite the low spectral resolution of the observations, the emission in all the tracers but DCO$^+$(2--1) is detected above 5$\sigma$ across 5 velocity channels (i.e. covering a width of about 18~km~s$^{-1}$), and therefore some kinematical information can already be extracted from these maps. We note, however, that in the case of DCN(2--1) this velocity spread is partly due to the hyperfine splitting (see Sect.~\ref{parameters} below). In Fig.~\ref{fcorespec} we show the average spectra within the \textit{main}, \textit{west}, and \textit{south} sources, defined by the polygons shown in Fig.~\ref{fint} (right panel). Figure~\ref{fvelo} presents the first moment maps for the four molecular tracers observed with the PdBI (colour scale), with overlaying contours of the corresponding velocity-integrated maps. The spectra, channel maps and velocity maps reflect the following properties :

\begin{table}[!hbt]
\caption{OMC-2~FIR~4 sources: masses and column densities\tablefootmark{a}} 
\label{tmass}     
\centering  
\begin{tabular}{lcccc}
\hline\hline 
Source & $M$(20~K) & $M$(50~K) & $N_\mathrm{H_2}$(20~K) & $N_\mathrm{H_2}$(50~K) \\ 
 & (M$_\odot$) & (M$_\odot$) & ($10^{23}$~cm$^{-2}$) & ($10^{23}$~cm$^{-2}$)\\
\hline   
Total & 25.7 & 9.2 & 19.2 & 6.9\\
\hline
Main & 9.1 & 3.2 & 29.0 & 10.4\\
West & 4.7 & 1.7 & 18.7 & 6.7\\
South & 3.0 & 1.1 & 12.4 & 4.4\\
\hline  
\end{tabular}\\
\tablefoottext{a}{Derived from the naturally-weighted 2-mm continuum map}
\end{table}

\begin{itemize}
\item The \textit{main} source displays a velocity gradient in the south-west to north-east direction.
\item The \textit{west} source as seen by CH$_3$OH and C$^{34}$S(3--2) appears slightly blue-shifted with respect to the systemic velocity of the source.
\item The DCO$^+$(2--1) emission is significantly the narrowest in line width and concentrates almost exclusively on the \textit{south} source.
\item A secondary, well-separated condensation is distinguished in the CH$_3$OH and DCN(2--1) maps to the south-west of the \textit{main} source, suggesting the presence of an additional physical source. The DCO$^{+}$ emission also extends towards this area.
\end{itemize}

 \begin{figure}[!htb]
   \centering
   \includegraphics[scale=0.5]{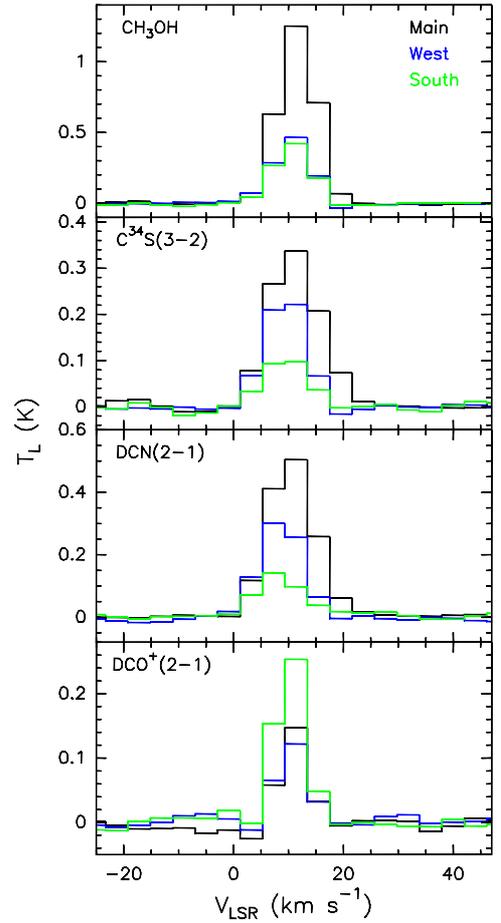}
      \caption{Spectra in the three main sources identified in OMC-2~FIR~4 (see text) for each line tracer. The vertical scale is the brightness temperature of the line, $T_\mathrm{L}$, averaged across the area of the corresponding polygon (see Fig.~\ref{fint}).}
         \label{fcorespec}
   \end{figure}

While the channel maps suggest the presence of more sources apart from the three we have identified so far, we cannot confirm it with 100\% certainty at this spatial resolution, and even less guess how many more are present. We therefore conclude, as stated above, that there are \textit{at least} three distinct sources within OMC-2~FIR~4, leaving open the possibility of further multiplicity within each of these three regions.

\subsubsection{Ammonia emission}\label{nh3}

\begin{figure}[!htb]
   \centering
   \includegraphics[scale=0.5]{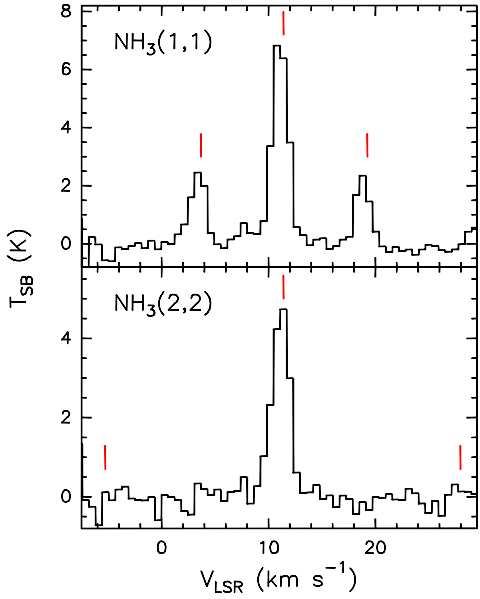}
      \caption{VLA ammonia spectra obtained towards the peak coordinates of OMC-2~FIR~4, with a channel width of 0.62~km~s$^{-1}$. The different hyperfine components are marked with red vertical lines.}
         \label{fnh3}
   \end{figure}

The NH$_3$(1,1) and (2,2) line spectra seen towards the peak NH$_3$(1,1) emission coordinates are shown in Fig.~\ref{fnh3}, with the different hyperfine components marked in red. The limited band width of the correlator allows us to cover only one pair of satellites for each transition, apart form the main component. In the case of NH$_3$(2,2), only the main component is detected. 
 
\begin{figure}[!hbt]
 \centering
 \includegraphics[scale=0.45]{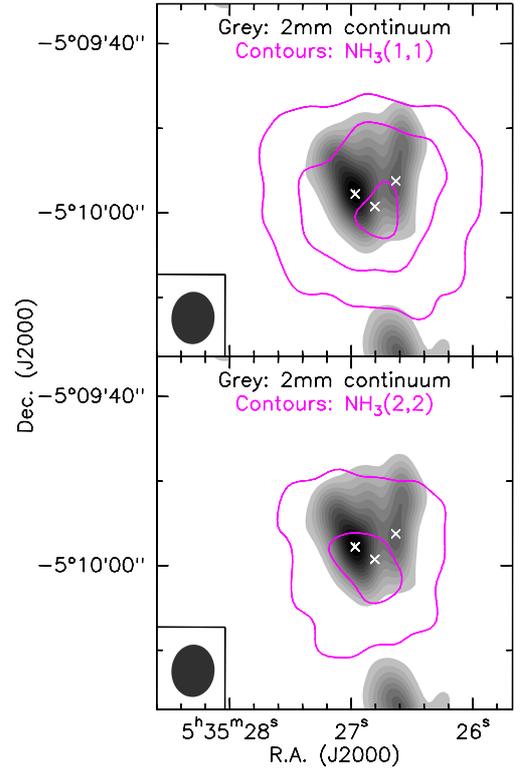}
 \caption{Velocity-integrated VLA ammonia contour maps obtained towards OMC-2~FIR~4, overlaid on the naturally weighted 2mm-continuum map (grey scale). Contours for all the maps start at 3$\sigma$ and increase by steps of 3$\sigma$. The 1$\sigma$ RMS values are 20~mJy~beam$^{-1}$~km~s$^{-1}$for both NH$_3$(1,1) and NH$_3$(2,2). White crosses mark the positions of the three sources identified in this work (see Table~\ref{tflux}).}
 \label{fnh3_int}
 \end{figure}

Velocity-integrated ammonia maps are shown in Fig.~\ref{fnh3_int}, where the emission corresponds to the main hyperfine component of the NH$_3$(1,1) and NH$_3$(2,2) transitions. These are overlaid on the 2-mm continuum PdBI image (grey scale) for comparison. We stress that the presence of extended emission which was filtered out by the VLA results in a relatively high 1$\sigma$ RMS level in the clean maps. Indeed, Li et al.~(\cite{li12}) recently published these ammonia data merged with single-dish maps obtained with the Green Bank Telescope (GBT), and a comparison with our VLA-only spectra indicates a flux loss of about 40\%. Nevertheless, in terms of morphology, the NH$_3$ emission is well defined around OMC-2~FIR~4 and peaks to the South of the \textit{main} source. The emission appears extended, with a roundish shape, and covers an area which is larger than that of the 2-mm continuum.
   
\begin{figure*}[!hbt]
   \centering
   \begin{tabular}{c}
   \includegraphics[angle=-90,scale=0.7]{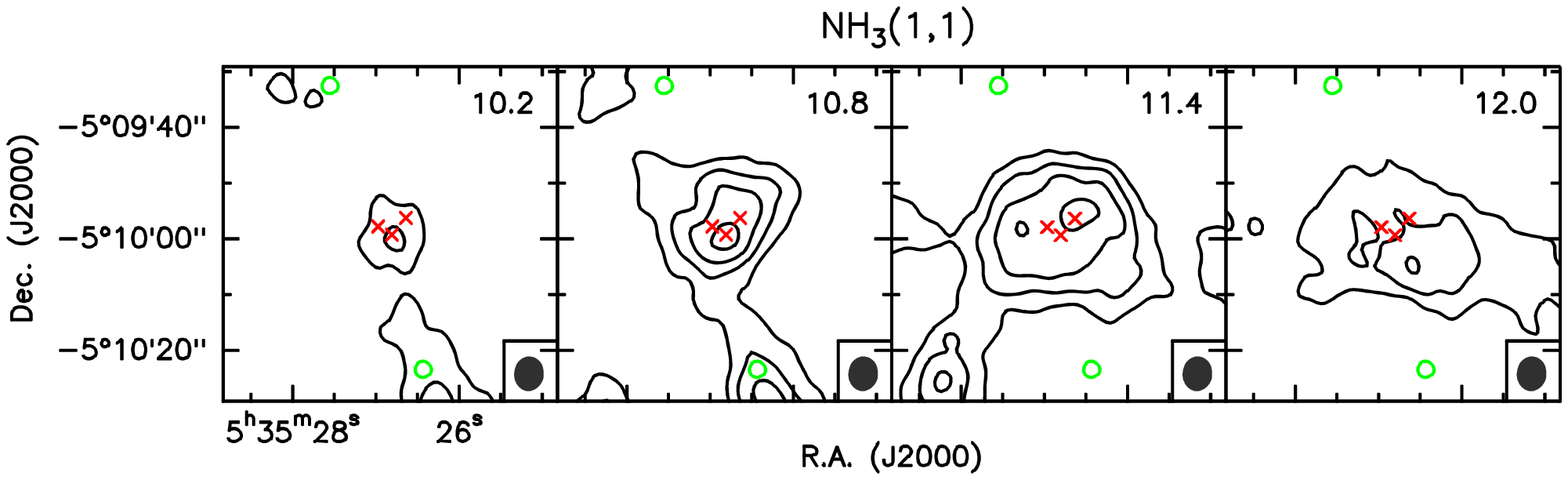}\\
   \includegraphics[angle=-90,scale=0.7]{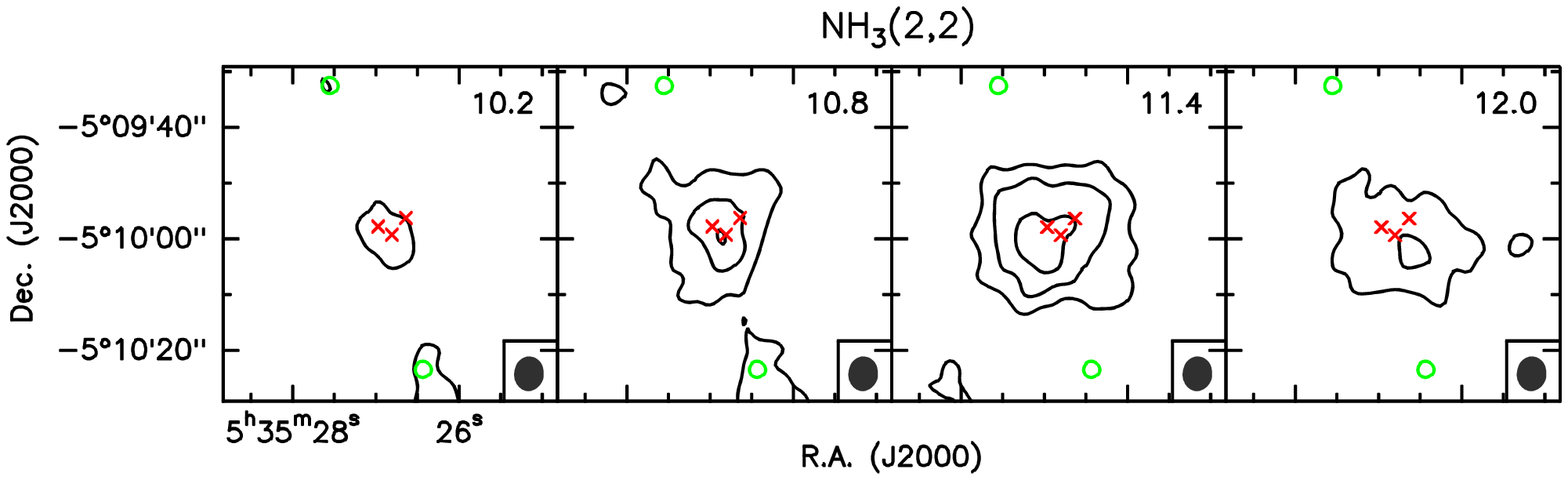}\\
   \end{tabular}
      \caption{VLA NH$_3$(1,1) and NH$_3$(2,2) channel maps towards OMC-2~FIR~4. Contours start at 5$\sigma$ and increase by steps of 5$\sigma$. The 1$\sigma$ RMS values are 5.0 and 4.7~mJy~beam$^{-1}$ for NH$_3$(1,1) and NH$_3$(2,2), respectively. The velocity of each channel is marked on the top-right corner of each panel in units of km~s$^{-1}$. Red crosses mark the positions of the three sources identified in this work (see Table~\ref{tflux}). Green circles mark the position of FIR~3 (NE) and FIR~5 (SW) according to the 1.2~mm observations of Chini et al.~(\cite{chini97}). The bottom-right ellipses in each panel represent the beam sizes.}
         \label{fnh3_ch}
   \end{figure*}

For completeness, in Fig.~\ref{fnh3_ch} we present the channel maps for the two NH$_3$ transitions, with the velocity of the channels given on the top-right corner of each panel in km~s$^{-1}$. Unlike the millimetre molecular spectra (Sect.~\ref{pdb}), the NH$_3$ lines are rather narrow, as can be seen from the velocity range covered by the maps (see also the spectra in Fig.~\ref{fnh3} and the $FWHM$ values in Table~\ref{tnh3fit}). This may partly be due to the better spectral resolution and lower sensitivity of these observations. Notice indeed that the channel spacing of the VLA observations ($\sim$0.6~km~s$^{-1}$) is substantially smaller than that of the WIDEX PdBI spectra.

The channel maps show extended emission around the central coordinates of OMC-2~FIR~4, and several other sources of emission: two towards the north-east and the south-west, where FIR~3 and FIR~5 are located, respectively (green circles), and a third one towards the south-east at the systemic velocity of the source (11.4~km~s$^{-1}$), also seen in the single-dish 1.3-mm maps carried out by Chini et al.~(\cite{chini97}). 

\subsubsection{Derivation of physical parameters}\label{parameters}

Both the DCN(2--1) and DCO$^+$(2--1) lines are split into six hyperfine components. For DCO$^{+}$(2--1), all of them lie within a velocity range of 0.3~km~s$^{-1}$, while for DCN(2--1) they span 8 km~s$^{-1}$. As a result, at the spectral resolution of our observations, they are all blended and unresolved, and therefore a hyperfine analysis is not possible. We have thus performed a Gaussian fit to each line and adopted a range of $T_\mathrm{ex}$ between 10 and 100~K to estimate a range of column densities from:

\begin{equation}
\label{enmol2}
N_\mathrm{mol} = \frac{8 \pi\ \nu^3}{c^3} \frac{Q}{A_\mathrm{ul}\ g_\mathrm{u}} \frac{e^{\frac{E_\mathrm{u}}{k T_\mathrm{ex}}}}{e^{\frac{h \nu}{k T_\mathrm{ex}}}-1} \frac{\int T_\mathrm{L}\ dv}{J(T_\mathrm{ex})-J(T_\mathrm{bg})}
\end{equation}
where $\nu$, $c$, $h$, $k$, $A_\mathrm{ul}$, $\tau_\mathrm{peak}$, $\Delta V$, $Q$, $g_\mathrm{u}$, $E_\mathrm{up}$ and $\int T_\mathrm{L} dv$ are, respectively, the frequency of the transition, the speed of light, Planck constant, Boltzmann constant, the spontaneous emission rate, the optical depth at the line intensity peak, the line $FWHM$, the partition function, the upper state degeneracy, the upper level energy and the velocity-integrated line intensity. This equation assumes optically thin conditions, and $J(T) = \frac{h \nu/k}{e^{h\nu/kT}-1}$, with $T_\mathrm{bg} = 2.7$~K. We have applied the same approach for the C$^{34}$S(3--2) transition.

The results of the Gaussian fits and the derived molecular column densities for each source are shown in Table~\ref{tabu}, where it can be seen that we are indeed dealing with optically thin conditions. Some of the rows in this Table probably correspond to ``artificial" sources, since not all the sources we have identified are well-defined or traced by all the lines, as explained above. We nevertheless present them for comparison and completeness. The  abundances of each molecule in the different sources are also listed. They have been obtained from the ratio of $N_\mathrm{mol}$ to the gas column density, $N$. For the latter, we have used the values listed in Table~\ref{tmass}, adopting both $T_\mathrm{d} = 20$~K and $T_\mathrm{d} = 50$~K.

\begin{table*}[!htb]
\caption{OMC-2~FIR~4 sources: Gaussian fit results, column densities and abundances for each PdBI tracer} 
\label{tabu}     
\centering  
\begin{tabular}{llcccccccc}
\hline\hline 
Tracer & Source & $T_\mathrm{L}$ & $V_\mathrm{LSR}$ & $FWHM$ & $T_\mathrm{ex}$\tablefootmark{a} & $\tau$ & $N_\mathrm{mol}$ & Abu. (20~K)\tablefootmark{b} & Abu. (50~K)\tablefootmark{b}\\
 &  & (K) & (km~s$^{-1}$) & (km~s$^{-1}$) & (K) &  & ($10^{12}$~cm$^{-2}$) & ($10^{-12}$) & ($10^{-12}$)\\
\hline
 & Main & 0.37 (0.01) & 10.9 (0.1) & 11.3 (0.2) & 10 - 100 & 0.055 - 0.004 & 15.3 - 38.2 & 5.3 - 13.2 & 14.7 - 36.7\\
C$^{34}$S(3--2) & West & 0.38 (0.01) & 9.3 (0.1) & 10.5 (0.2) & 10 - 100 & 0.040 - 0.003 & 8.8 - 22.2 & 4.7 - 11.9 & 13.1 - 33.1\\
 & South & 0.15 (0.01) & 9.3 (0.2) & 10.7 (0.6) & 10 - 100 & 0.017 - 0.001 & 4.0 - 10.3 & 3.2 - 8.3 & 9.1 - 23.4\\
\hline
 & Main & 0.55 (0.01) & 10.6 (0.1) & 10.0 (0.1) & 10 - 100 & 0.085 - 0.006 & 6.8 - 22.0 & 2.3 - 7.6 & 6.5 - 21.2\\
DCN(2--1) & West & 0.67 (0.01) & 9.8 (0.1) & 9.8 (0.2) & 10 - 100 & 0.052 - 0.003 & 3.7 - 12.5 & 2.0 - 6.7 & 5.5 - 18.7\\
 & South & 0.29 (0.01) & 9.0 (0.1) & 10.2 (0.4) & 10 - 100 & 0.022 - 0.001 & 1.8 - 6.0 & 1.5 - 4.8 & 4.1 - 13.6\\
\hline
 & Main & 0.19 (0.01) & 10.8 (0.2) & 6.1 (0.3) & 10 - 100 & 0.024 - 0.002 & 0.6 - 2.3 & 0.2 - 0.8 & 0.6 - 2.2\\
DCO$^+$(2--1) & West & 0.20 (0.01) & 11.1 (0.2) & 6.3 (0.3) & 10 - 100 & 0.020 - 0.001 & 0.6 - 2.1 & 0.3 - 1.1 & 0.9 - 3.1\\
 & South & 0.38 (0.01) & 10.3 (0.1) & 6.3 (0.2) & 10 - 100 & 0.044 - 0.003 & 1.3 - 4.5 & 1.0 - 3.6 & 3.0 - 10.2\\
\hline
\end{tabular}\\
\tablefoottext{a} Adopted values\\
\tablefoottext{b} Abundance determined using the dust temperature indicated in brackets to derive the total gas column density (see Table~\ref{tmass})\\
\end{table*}

As for the methanol lines, we analysed them by means of a Large Velocity Gradient (LVG) code, originally developed by Ceccarelli et al.~(\cite{cec03}) and modified to include the methanol collisional coefficients with H$_2$ by Pottage et al.~(\cite{pot04}). For this analysis, we used the methanol lines obtained by averaging the emission over the entire OMC-2~FIR~4 clump. The line intensities thus measured are listed in Table~\ref{tlvg}, where the values shown for the 145-GHz lines do not correspond to single transitions but to ensembles of blended methanol lines around the indicated frequency and transition. We ran a large grid of models to cover a wide parameter space: N(CH$_3$OH) from $10^{13}$ to $4\times 10^{15}$ cm$^{-2}$, density from $10^4$ to $10^9$ cm$^{-3}$ and temperature from 10 to 200 K. The observed lines were compared with the LVG model predictions and the best fit solution was found minimising the $\chi^2$ value with respect to the three above parameters, assuming a source size of $15'' \times 15''$, as measured in our maps for the whole OMC-2~FIR~4 clump. We note that for the blended lines the predicted intensities were obtained by adding the intensity by each single line in the velocity interval.

We found that the CH$_3$OH column density is very well constrained between 0.6 and 1 $\times 10^{15}$ cm$^{-2}$, roughly consistent with the value of $\sim 2 \times 10^{14}$~cm$^{-2}$ reported by Kama et al.~(\cite{kama10}) from the early results of the HIFI data. The density is between 0.2 and 1 $\times 10^7$ cm$^{-3}$, a range of values that matches the average density we obtain from the millimetre continuum emission (see Sect.~\ref{pdb}). The predicted optical depths of the CH$_3$OH lines are low, around a few times $10^{-2}$.

The observed lines do not allow to constrain the temperature, except that it is larger than 20 K. The larger the CH$_3$OH column density the larger the gas temperature. For example, if N(CH$_3$OH) is $8 \times 10^{14}$ cm$^{-2}$ then the gas temperature is larger than 50 K; for N(CH$_3$OH)~$= 1 \times 10^{15}$ cm$^{-2}$ the gas temperature is larger than 120 K. More methanol transitions will be necessary to better constrain the temperature, but our analysis already shows that the gas emitting the methanol lines must be relatively dense. The best fit solutions of the LVG modelling have a reduced $\chi^2$ below 0.5 and are summarised in Table~\ref{tlvg}.

\begin{table}[!htb]
\caption{CH$_3$OH line intensities and results of the LVG analysis} 
\label{tlvg}     
\centering  
\begin{tabular}{cccc}
\hline \hline
\multicolumn{4}{c}{CH$_3$OH line intensities}\\
\hline
Transition & $\nu$ & $\int T_\mathrm{SB} dV$ & Notes\\
 & (GHz) & (K~km~s$^{-1}$) & \\
\hline
$3_{1,3} - 2_{1,2}$~A$^+$ & 143.866 & $5.2 \pm 1.6$ & isolated\\
$3_{-1,3} - 2_{-1,2}$~E$^+$ & 145.097 & $28.8 \pm 8.7$ & 3 blended lines\\
$3_{-2,2} - 2_{-2,1}$~E$^+$ & 145.126 & $13.9 \pm 4.2$ & 5 blended lines\\
\hline\hline 
\multicolumn{4}{c}{Results of the LVG analysis}\\
\hline
N(CH$_3$OH) & Density & Temperature & \\
(10$^{14}$~cm$^{-2}$) & (10$^6$~cm$^{-3}$) & (K) & \\
\hline
6 & $> 2$ & 20 - 50\\
8 & $> 2$ & $> 50$\\
10 & 2 - 10 & $> 120$\\
\hline
\end{tabular}
\end{table}

Finally, we performed a hyperfine line fit to the two ammonia spectra in Fig.~\ref{fnh3} using the standard NH$_3$(1,1) and NH$_3$(2,2) methods in CLASS. We subsequently derived the corresponding rotational and kinetic temperatures, as well as the NH$_3$ column density, following the steps summarised in Busquet et al.~(\cite{gemma09}) and assuming a beam filling factor equal to 1. The results are reported in Table~\ref{tnh3fit}, where the peak synthesised beam temperature, $T_\mathrm{SB}$, the velocity, $V_\mathrm{LSR}$, and the optical depth, $\tau_\mathrm{main}$ refer to the main component of the transition. We note that the errors on the temperature and column density estimates are of the order of or greater than the values themselves, mostly due to the large errors on $T_\mathrm{SB}$ and $\tau$. Using the H$_2$ column densities listed in Table~\ref{tmass}, we obtain an NH$_3$ abundance slightly above 10$^{-10}$. While this value is about 2 orders of magnitude smaller than what is typically reported in the literature, there are studies of ammonia in dense molecular cores that find values as low as ours or even lower (e.g. Foster et al.~\cite{foster09}, Wootten~\cite{woo95}).

\begin{table*}[!tbh]
\caption{Results of the line fitting to the NH$_3$(1,1) and (2,2) lines and derived physical parameters} 
\label{tnh3fit}     
\centering  
\begin{tabular}{cccccc}
\hline \hline
\multicolumn{6}{c}{Results of the hyperfine line fitting}\\
\hline
Line & $T_\mathrm{SB}$ & $V_\mathrm{LSR}$ & $\Delta V$ & $\tau_\mathrm{main}$ \\ 
 & (K) & (km~s$^{-1}$) & (km~s$^{-1}$) & \\ 
\hline   
NH$_3$(1,1) & 7.5 (2.7) & 11.1 (0.1) & 1.5 (0.1) & 0.7 (0.2)\\
NH$_3$(2,2) & 4.9 (9.8) & 11.2 (0.1) & 1.9 (0.1) & 0.1 (0.2)\\
\hline \hline
\multicolumn{6}{c}{Physical parameters}\\
\hline
$T_\mathrm{ex}$ & $T_\mathrm{rot}$ & $T_\mathrm{kin}$ & $N_\mathrm{NH_3}$ & Abu. (20~K)\tablefootmark{a} & Abu. (50~K)\tablefootmark{a}\\
(K) & (K) & (K) & ($10^{14}$~cm$^{-2}$) & ($10^{-10}$) & ($10^{-10}$)\\
\hline
18.2 & 22.7 & 29.6 & 2.6 & 1.4 & 3.8\\
\hline
\end{tabular}\\
\tablefoottext{a} Abundance determined using the dust temperature indicated in brackets to derive the total gas column density (see Table~\ref{tmass})\\
\end{table*}

\section{Discussion}\label{discuss}

\subsection{Source multiplicity within OMC-2~FIR~4}

Several studies (J\o rgensen et al~\cite{jor06}, S08, Crimier et al.~\cite{crim09}, Kama et al.~\cite{kama10}) suggest that OMC-2~FIR~4 contains three main components:

\begin{itemize}
\item A large-scale cool envelope about 10000~AU (i.e. 0.05~pc) across.
\item An inner hot core, with kinetic temperatures above 100~K and around 400~AU in size.
\item A shock region produced by the interaction of an external bipolar outflow driven by the nearby OMC-2~FIR~3, which lies to the north-east of FIR~4.
\end{itemize}

In addition, S08 report the presence of 11 cores seen at 3-mm inside this region, and speculate that such fragmentation has been externally triggered by the FIR~3 outflow.

The emission of all the maps presented in Sect.~\ref{results} covers roughly 10000~AU (0.05~pc) and therefore traces the envelope component of OMC-2~FIR~4. In particular, this component appears to be better traced by the NH$_3$ maps, which are essentially single-peaked and display the most extended emission, similar in size and morphology to the structures detected in previous single-dish observations (e.g.~Chini et al.~\cite{chini97}, the ASTE maps of Shimajiri et al.~\cite{shima08}). The temperature derived for the ammonia emission, $\sim 30$~K, is typical of protostellar (in our case protocluster) envelopes (e.g. Olmi et al.~\cite{olmi12}).

At the resolution of our PdBI observations, despite being slightly better than that of the NMA observations, we cannot confirm the existence of the 11 cores claimed by S08. Inspection of the different maps presented in this work allows us to clearly separate three regions, denoted by the identifiers \textit{main}, \textit{west}, and \textit{south} (Sect.~\ref{pdb}). It appears, however, that our 2-mm continuum and some molecular line maps indeed hint at the presence of multiple smaller condensations within the three regions we detect, but these will need to be confirmed by higher angular resolution observations.

At this point, it is worth mentioning that the hot core hypothesis proposed by Crimier et al.~(\cite{crim09}) is based on a single source at the centre of the large-scale envelope. S08 and our new observations show that the envelope component breaks down at scales of $\sim 2000$~AU, much larger than the presumed hot core size (with a radius of 440~AU). Only high-angular resolution imaging of typical hot core tracers (e.g. complex organic molecules) can observationally prove the existence of a hot core.

As shown in Fig.~\ref{fvelo}, a velocity difference of about 6~km~s$^{-1}$ can be seen between the south-west and the north-east extremes of the \textit{main} source in the CH$_3$OH and C$^{34}$S(3-2) maps, with the north-eastern tip being the most red-shifted. S08 found the same behaviour in their single-dish methanol maps. Additional interferometric observations of CO and SiO lines provided the same authors with morphological and dynamical evidence which led them to suggest that there is an interaction between the north-eastern edge of OMC-2~FIR~4 and the molecular outflow driven by the nearby FIR~3. The red-shifted emission we detect from CH$_3$OH and C$^{34}$S(3--2) may also be explained by this scenario, i.e., it arises from shocked gas from the interaction between the FIR~3 outflow and the dense gas of FIR~4. It might alternatively be tracing the FIR~3 outflow itself, which cannot be detected much beyond the boundary of the FIR~4 clump due to the limited sensitivity of the PdBI to more extended structures. More observations are necessary to confirm the external outflow interpretation, since with our data alone we are not able to conclude that this is indeed the case. Other possibilities exist that could cause a velocity gradient like the one we detect, such as rotation or internal outflowing/expanding motions.

\subsection{Nature of the sources}\label{nature}

Our observations allow us to distinguish three separate sources which may or may not present further substructure:

\begin{enumerate}
\item The \textit{main} source, at the nominal coordinates and systemic velocity (11.4~km~s$^{-1}$) of OMC-2~FIR~4. It is the strongest 2-mm continuum and methanol source in the region. Its elongated shape suggests that it harbours two or more individual cores, although this hypothesis can only be confirmed with higher spatial resolution imaging. The red-shifted velocity of its northern portion might indicate an interaction with the red-shifted lobe of the external outflow driven by FIR~3.
\item The \textit{west} source, with a systemic velocity of 9-10~km~s$^{-1}$. It is well traced by C$^{34}$S(3--2) and DCN(2--1), and weaker than the \textit{main} source emission.
\item The \textit{south} source, also with $V_\mathrm{LSR} = 9$-10~km~s$^{-1}$, best traced by DCO$^+$(2--1). It is the faintest methanol, DCN, and 2-mm continuum source, and has no significant emission in C$^{34}$S(3--2). Its association with DCO$^+$ indicates this region is likely colder than the other two, suggesting that it contains either a starless core or a protostar in an earlier evolutionary stage or with a lower mass (e.g. J\o rgensen et al.~\cite{jor04}). Alternatively, since it does not display any emission peak in the 2-mm continuum image, it may not be a separated molecular core, but part of the cold external envelope of OMC-2~FIR~4.
\end{enumerate}

From the results of our LVG analysis on the CH$_3$OH lines, we find that the entire OMC-2~FIR~4 clump is rather dense ($n > 2 \times 10^6$~cm$^{-3}$ at 15$''$, i.e. 6000~AU). This suggests that the three regions we identify in this work, which are altogether smaller in size when compared to the whole emission considered in the CH$_3$OH analysis, may be even denser. We note that the density we derive is in good agreement with S08 and with the predictions made by Crimier et al.~(\cite{crim09}), both of which report a value of $10^6$~cm$^{-3}$.

\subsection{An ionised region in OMC-2~FIR~4}\label{hii}

\begin{figure*}[!hbt]
\centering
\includegraphics[scale=0.5]{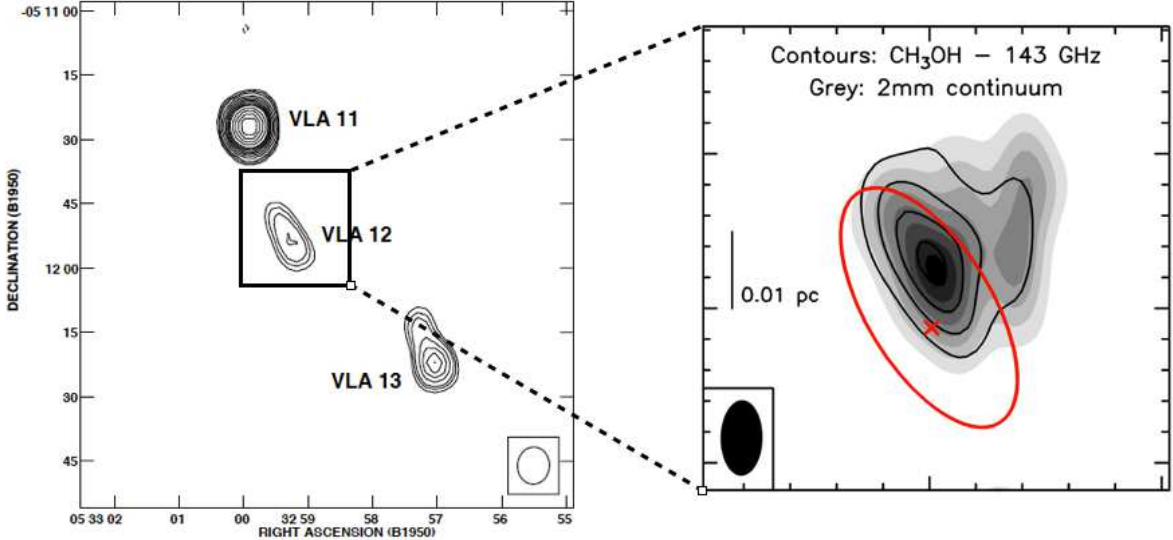}
\caption{\textit{Left panel}: VLA map of the 3.6~cm continuum emission in the OMC-2~FIR~4 region and its surroundings (taken from Reipurth et al.\ 1999). The synthesised beam is shown at the bottom-right corner. The source labelled VLA~12 coincides with the position of OMC-2~FIR~4. \textit{Right panel}: Close-up view showing our PdBI continuum (grey scale) and CH$_3$OH (black contours) maps. The synthesised beam is shown at the bottom-left corner. The red ellipse schematically represents the 4$\sigma$ contour of the VLA image (left panel). The red cross marks the position of the cm emission peak.}
\label{fvla}
\end{figure*}

Radio-continuum observations of the OMC2/3 filament were carried out with the VLA by Reipurth et al.~(\cite{rei99}) at 3.6~cm. At the position of OMC-2~FIR~4, they detected an elongated source which the authors interpret as a radio jet. When comparing their map with our PdBI map, despite the different angular resolutions (i.e. $\sim 4''$ for the PdBI maps, $8''$ for the VLA maps), we find that the position and shape of the radio source coincides fairly well within the pointing errors of the two maps, with the emission distribution of our 2-mm continuum and methanol PdBI images, which are both elongated in the same direction as the VLA source (see Fig.~\ref{fvla}).

While Reipurth et al.~(\cite{rei99}) attribute the radio continuum emission to a jet, the interferometric observations performed towards this source so far (e.g. S08, this work) have failed to detect molecular line emission at high velocities indicative of a jet or outflow emanating from this region. This leads us to consider a different possibility, namely that the radio emission arises from a very compact gas cavity ionised by a central young star, i.e. an H{\sc ii} region, which is either surrounded by a shell of dust and molecular gas best traced by the 2-mm continuum and CH$_3$OH emission, or else harbours two or more unresolved molecular dense cores within it. The H{\sc ii} region scenario is a plausible explanation for the continuum radio emission, since IM protostars can also produce sufficiently intense H{\sc ii} regions to be easily detected at centimetre wavelengths (e.g. Thompson~\cite{thom84}).

Reipurth et al.~(\cite{rei99}) measured a flux of 0.64~mJy at 3.6~cm towards OMC-2~FIR~4. In the scenario of an optically thin H{\sc ii} region, and assuming a single zero-age main sequence star is powering it, this implies a B3-B4 spectral type for the central star (using the tables of e.g. Panagia~\cite{pan73}, Thompson~\cite{thom84}, Martins et al.~\cite{mar05}). This translates into a luminosity between 700 and 1000~L$_\odot$ and would indicate that this young star is the most massive source in the protocluster. It is also possible, given the elongated shape of the radio emission, that two (or more) individuals are responsible for the ionisation of their surroundings. Only higher angular resolution and sensitivity images of the continuum emission at different wavelengths will allow us to verify this possibility and find out whether the nature of the centimetre emission is an H{\sc ii} region, as proposed here, or a radio jet.

\subsection{A final note on luminosities}

There is a certain confusion regarding the luminosity of OMC-2~FIR~4 in the literature. While the first estimates by Mezger et al.~(\cite{mez90}), and later by Crimier et al.~(\cite{crim09}), report a value between 400~L$_\odot$ and 1000~L$_\odot$, a recent work by Adams et al.~(\cite{adams12}) yields 50~L$_\odot$. In what follows we try to understand the reason of this discrepancy.

The first point to mention is the complexity of the region. Indeed, the morphology and size of OMC-2~FIR~4 change when seen at different wavelengths: in the mid-IR it appears to be a compact source, while at wavelengths above $\sim 100$~$\mu$m its emission is more extended ($\sim 30''$ in diameter) and peaks around $5''$ west of the mid-IR source. This said, we note that the luminosity estimates published by the different authors refer to different components. On the one hand, Adams et al.~(\cite{adams12}) built the Spectral Energy Distribution of the source by measuring the continuum fluxes, at IR and sub-mm wavelengths, of the compact source they detect at 19~$\mu$m and 37~$\mu$m with SOFIA. In other words they adopted small aperture sizes, ranging between 7$''$ and 13$''$ and centred at the mid-IR peak. On the other hand, Crimier et al.~(\cite{crim09}) and Mezger et al.~(\cite{mez90}) took into account a larger area, of 30$''$ and 50$''$ in diameter, respectively, which includes the whole envelope of OMC-2~FIR~4. In particular:

\begin{itemize}
\item Crimier et al.~(\cite{crim09}) used far-IR data from ISO and IRAS, whose coarse angular resolution likely led to measured fluxes that were overestimated, due to possible contamination from the nearby protostar FIR~3, by at most a factor 2, since at such wavelengths the dominant emission source is FIR~4.
\item Adams et al.~(\cite{adams12}) obtained \textit{Herschel}-PACS maps of OMC-2~FIR~4 from which they measured the fluxes at 70~$\mu$m and 160~$\mu$m. They did so by applying a small aperture of 9.6$''$ and 12.8$''$, respectively, around the mid-IR peak coordinates, and a background annulus that actually covers part of the extended emission of the region. In addition, they carried out APEX observations at 350~$\mu$m and 850~$\mu$m which lacked the sufficient sensitivity to detect any emission. They therefore measured an upper limit of the flux considering only the emission within a beam size (7.3$''$ and 19$''$, respectively), again around the mid-IR coordinates, and therefore offset by $\sim 5''$ from the actual sub-mm peak position. As a result, their reported sub-mm fluxes are respectively 10 and 3 times lower than those detected and measured by Crimier et al.~(\cite{crim09}) with the CSO and JCMT telescopes. As a result, considering these points, the authors may have underestimated the flux of OMC-2~FIR~4 by a factor of up to 10.
\end{itemize}

\begin{table}[!h]
\caption{Mass and luminosity estimates for OMC-2~FIR~4\tablefootmark{a}} 
\label{tlum}     
\centering  
\begin{tabular}{cccl}
\hline\hline 
$L$ & $M$ & Size & Ref.\\
(L$_\odot$) & (M$_\odot$) & ($''$) & \\ 
\hline
350 & 34 & 50 & Mezger et al.~(\cite{mez90})\\
920 & 30 & 30 & Crimier et al.~(\cite{crim09})\\
50 & 10 & 8 & Adams et al.~(\cite{adams12})\\
\hline
\end{tabular}
\\
\tablefoottext{a}{Values re-scaled to a distance of 420~pc}
\end{table}

In short, it appears that the discrepancy in luminosity estimates found in the literature is caused by the fact that different authors measure different components of OMC-2~FIR~4, with Adams et al~(\cite{adams12}) concentrating on the compact mid-IR source and Crimier et al.~(\cite{crim09}) on the entire envelope, which is several times larger. Table~\ref{tlum} summarises the different values for the luminosity, mass and size of OMC-2~FIR~4 reported in the literature. Based on the discussion above, the actual value depends on how one defines OMC-2~FIR~4 and its boundaries. If our interpretation of an H{\sc ii} region is correct (see Sect.~\ref{hii}), the luminosity of the whole protocluster could be around several hundred solar luminosities.

\section{Conclusions}\label{conclusions}

\begin{figure}[!hbt]
\centering
\includegraphics[scale=0.4]{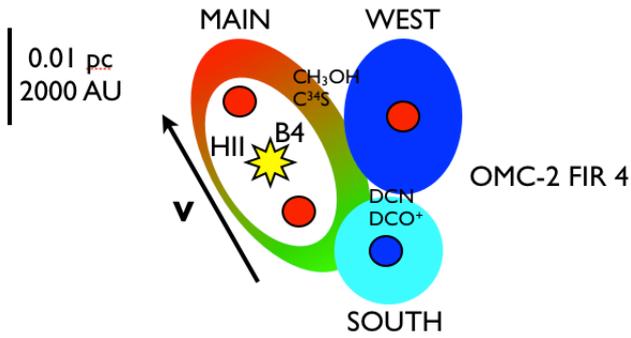}
\caption{Schematic cartoon of OMC-2~FIR~4 and the nearby OMC-2~FIR~3, summarising all the physical elements in the region that we propose according to the data presented in this work (see Sect.~\ref{discuss}). The white ellipse inside the \textit{main} source depicts the ionised region powered by a B4 young star. The colours of the \textit{main}, \textit{west}, and \textit{south} sources represent their systemic velocities with respect to the nominal value for OMC-2~FIR~4 ($V_\mathrm{LSR} = 11.4$~km~s$^{-1}$), marked in green. The small red and blue circles within the three sources represent the possibility that they harbour smaller unresolved molecular condensations, with the colour blue denoting a colder temperature than red.}
\label{fcartoon}
\end{figure}

We have presented new interferometric observations carried out towards the intermediate-mass protocluster OMC-2~FIR~4, in the Orion~A complex. These include PdBI maps of the continuum emission at 143.4~GHz (2~mm) and the CH$_3$OH(3,1,+0 -- 2,1,+0), C$^{34}$S(3--2), DCN(2--1), and DCO$^{+}$(2--1) lines, as well as VLA maps of the NH$_3$(1,1) and (2,2) inversion transitions. Despite the coarse channel spacing of the observations (0.6~km~s$^{-1}$ for the VLA spectra and 4~km~s$^{-1}$ for the PdBI images), we are able to extract important information about the morphology and kinematics of the region. Figure~\ref{fcartoon} shows a cartoon of the OMC-2~FIR~4/FIR~3 region, summarising all the elements present according to our interpretation of the data presented in this work. Our findings are summarised as follows:

   \begin{enumerate}
      \item Our 2-mm continuum maps reveal two main components and hints of secondary cores or clumps surrounding it. The existence of the 11 cores discovered by S08 cannot be confirmed by our observations, despite our higher angular resolution.
      \item Each PdBI line traces the region differently. By comparing their emission, we are able to distinguish three separate sources of one or several solar masses each, which we have named \textit{main}, \textit{west} and \textit{south}. The \textit{main} and \textit{west} ones are warmer. The \textit{south} source displays a stronger line emission in DCO$^{+}(2-1)$, suggesting a colder temperature and therefore an earlier evolutionary state or lower mass core. Alternatively, it may be associatied with part of the colder envelope of OMC-2~FIR~4. As for the ammonia maps, their emission is more extended and essentially single-peaked, and therefore traces the envelope of the protocluster. An LVG modelling of the CH$_3$OH lines we observe allows us to conclude that the whole protocluster is very dense, with $n \gtrsim 2 \times 10^6$~cm$^{-3}$, in agreement with previous theoretical predictions on the structure of the envelope of OMC-2~FIR~4. 
      \item The \textit{main} source as seen by the 2-mm continuum and CH$_3$OH emission coincides, both in position (within the pointing errors) and in shape, with the radio-continuum emission observed at 3.6~cm with the VLA (Reipurth et al.~\cite{rei99}). In absence of clear interferometric evidence of a molecular outflow driven by FIR~4, we interpret this as an H{\sc ii} region powered either by a B3-B4 type young star, or by two or more lower-mass stellar objects. This region is likely surrounded by a shell of dust and molecular gas (seen by the millimetre continuum and CH$_3$OH maps). This scenario remains to be verified by higher angular resolution and multi-wavelength observations.
   \end{enumerate}
   
Due to its relative proximity to the Sun, OMC-2~FIR~4 is an ideal laboratory to explore intermediate- and low-mass star formation in the context of a whole protocluster. The evidence of core multiplicity in this region, coupled with the insufficient spatial resolution of our observations to clearly separate them, highlights the need of higher spectral and angular resolution imaging to disentangle the different components and extract more information about the kinematics and the excitation conditions in each of them. This information will be of crucial importance to interpret the \textit{Herschel}-HIFI data obtained towards this source within the CHESS Key Programme.

\begin{acknowledgements}
      A.L.S. and C.C. acknowledge funding from the CNES (Centre National d'\'Etudes Spatiales) and from the Agence Nationale pour la Recherche (ANR), France (project FORCOMS, contracts ANR-08-BLAN-022). M.K. gratefully acknowledges funding from an NWO grant, NOVA, Leids Kerkhoven-Bosscha Fonds and the COST Action on Astrochemistry. C.D. acknowledges funding from Leids Kerkhoven-Bosscha Fonds. A.F. has been partially supported within the program CONSOLIDER INGENIO 2010, under grant  CSD2009-00038 ``Molecular Astrophysics: The Herschel andALMA Era -- ASTROMOL".
\end{acknowledgements}

\end{document}